\documentclass[conference,a4paper]{IEEEtran}

\usepackage{epsfig,makeidx,color,mathtools,bigints}
\usepackage{graphicx}
\usepackage{amsbsy}
\usepackage{amsmath}
\usepackage{amssymb}
\usepackage{euscript,verbatim}

\newcommand{\R}{\mathbf{R}} 
\newcommand{\C}{\mathbf{C}} 

\newcommand{\diag}{{\hbox{diag}}}
\renewcommand{\det}{{\hbox{\rm det}}}

\begin{document}

\title{Multi-Dimensional and Non-Uniform Constellation Optimization via the Special Orthogonal Group}

\author{
  \IEEEauthorblockN{David A. Karpuk, Camilla Hollanti}
  \IEEEauthorblockA{Dept. of Mathematics and Systems Analysis\\
    Aalto University\\
    P.O.\ Box 11100\\
    FI-00076 Aalto, Finland\\
    emails: \{david.karpuk, camilla.hollanti\}@aalto.fi} 
}

\maketitle

\begin{abstract}

With the goal of optimizing the CM capacity of a finite constellation over a Rayleigh fading channel, we use one-parameter subgroups of the Lie group of rotation matrices to construct families of rotation matrices which optimize a certain objective function controlling the CM capacity.  Our construction does not depend on any assumptions about the constellation or signal-to-noise ratio.  We confirm the benefits of our construction for uniform and non-uniform constellations at a large range of SNR values through numerous simulations.  We show that in two and four dimensions one can obtain a further potential increase in CM capacity by jointly considering non-uniform and rotated constellations.

\end{abstract}


\IEEEpeerreviewmaketitle

\section{Introduction}


We consider the general problem of constructing good constellations for Rayleigh fast fading channels.  As in \cite{OV}, by using a bit interleaver, assuming perfect channel state information at the receiver, and separating real and imaginary parts, we can model the channel as
\begin{equation}\label{model}
y = Hx + z
\end{equation}
where 
\begin{itemize}
\item[$\bullet$] $x$ is the transmitted codeword, selected from a finite constellation $\mathcal{X}\subset \R^n$,
\item[$\bullet$] $H=\diag(\alpha_i)$ is a real diagonal $n\times n$ matrix with $\alpha_i$ a Rayleigh distributed random variable with $\mathbf{E}(\alpha_i^2) = 1$,
\item[$\bullet$] $z=(z_i) \in\R^n$ a noise vector with $z_i$ a real zero-mean Gaussian random variable with variance $N_0/2$, and 
\item[$\bullet$] $y\in \R^n$ the received vector.
\end{itemize}

Constellations for Rayleigh fading channels traditionally consist of uniformly spaced $M$-QAM symbols, i.e.\ a finite constellation $\mathcal{X}\subset \R^2$ of size $M=2^q$.  However, the DVB consortium has established rotated QAM constellations as a part of the DVB-T2 industry standard \cite{DVBT2}, and non-uniform QAM constellations have been considered in DVB-NGH (next-generation handheld) implementation \cite{DVBNGH} due to the improvement in CM (coded modulation) capacity.  A natural question is whether combining non-uniformity and rotations can offer an additional increase in capacity, and thus one would like to know how to find good rotations for arbitrary constellations, not just traditional $M$-QAM.

As is described in, for example, \cite{herath} and \cite{thaiBICM}, rotated multi-dimensional constellations have the potential to increase CM capacity at low- and mid-range SNR.  While the minimum product distance and resulting number theoretic constructions of \cite{OV} are good at reducing the pairwise error probability (PEP), these are based on asymptotic design criteria which may not be valid in the low SNR regime.  Taking into considering both non-uniformity and multi-dimensionality, we seek to construct rotations which maximize CM capacity for arbitrary constellations in arbitrarily large dimensions, which can be further optimized according to a given SNR.

Constructing good rotations in arbitrarily high dimensions means abandoning explicit parameterizations of such matrices, since such parameterizations become non-canonical and unwieldy as the dimension of the ambient space increases.  The collection of all rotations of $\R^n$ is the \emph{special orthogonal group} $SO(n)$ which has dimension $n(n-1)/2$ as a real manifold, meaning any parameterization of $n\times n$ rotation matrices requires at least $n(n-1)/2$ variables.

Our approach to constructing optimal rotation matrices differs largely from previous work on the subject, in that we abandon explicit parameterizations of rotation matrices in favor of the matrix exponential map
\begin{equation*}
\exp: \frak{so}(n) \rightarrow SO(n)
\end{equation*}
where $\frak{so}(n)$ is the Lie algebra of all skew-symmetric matrices.  The general mathematical framework of Lie groups and Lie algebras allows us to construct well-performing families $Q_{2^k}(t)$ of $2^k$-dimensional rotation matrices for all $k$, which we view as one-parameter subgroups of $SO(2^k)$.  The problem of optimizing over all $n(n-1)/2$ parameters defining a rotation matrix is thus reduced to optimizing over just a single parameter $t\in[0,2\pi]$, which is easily done by exhaustive search and can be catered to the given SNR.


\section{Related Work}

The idea of rotating two-dimensional constellations to obtain an increase in diversity was first presented in \cite{boulle}, and numerous algebraic and number theoretic techniques also exist to construct fully-diverse lattices with good minimum product distance \cite{OV}.  The current authors previously used numerical techniques on $SO(n)$ to construct rotations which attempt to minimise the pairwise error probability \cite{karpukrot}.

Our work is partially inspired by \cite{herath}, in which the authors constructed good rotation matrices for $4$-QAM and $16$-QAM constellations in $\C^4$ and $\C^6$ with the goal of optimizing capacity.  Complex multi-dimensional rotations have been used in \cite{thaiBICM} to increase the performance of BICM-ID systems for Rayleigh fading channels.  Furthermore, two-dimensional rotations have been considered in \cite{chineseBICM}, \cite{indianBICM} to improve BICM capacity, and in \cite{chineseLDPC}, \cite{dutchLDPC} in conjunction with LDPC codes.



\section{CM Capacity of Constellations in $\R^n$}\label{cap2d}

In this section we recall some familiar formulas for the CM capacity of an $n$-dimensional constellation $\mathcal{X}\subset\R^n$ of size $M$.  For the AWGN channel (\ref{model}), the mutual information of the output $\mathcal{Y}$ and the constellation $\mathcal{X}$ is
\begin{eqnarray*}
I^{\text{A}}(\mathcal{Y},\mathcal{X}) \hspace{-.75em}&=& \hspace{-.75em}\log_2(M) - \sum_{i = 1}^M \bigintssss_{\R^n}\frac{\exp\left(-||y-x_i||^2/N_0\right)}{M(\pi N_0)^{\frac{n}{2}}} \\
\hspace{-1em}&\times&\hspace{-.5em} \log_2\left[\sum_{k = 1}^M\exp\left(\frac{||y-x_i||^2-||y-x_k||^2}{N_0}\right)\right]\ dy
\end{eqnarray*}
If $H = \text{diag}(\alpha_i)$ is a fixed fading matrix, then the conditional mutual information of $\mathcal{Y}$ and $\mathcal{X}$ given $H$ is 
\begin{equation*}
I(\mathcal{Y},\mathcal{X};H) = I^{\text{A}}(\mathcal{Y},H\mathcal{X}),
\end{equation*}
where $H\mathcal{X}=\{Hx\ |\ x\in\mathcal{X}\}$.  We can define the \emph{coded modulation capacity} $\mathcal{X}$ over our Rayleigh fading channel by taking the expectation over all $H$:
\begin{equation*}
C^{\text{CM}}(\mathcal{X}) = \mathbf{E}_H I(\mathcal{Y},\mathcal{X};H)
\end{equation*}
If $Q$ is an $n\times n$ rotation matrix we let $C^{\text{CM}}(\mathcal{X},Q)$ denote the value of $C^{\text{CM}}(\mathcal{X})$ after $\mathcal{X}$ has been rotated by $Q$.  Our ultimate goal is to compute the rotation $Q$ which maximizes $C^{\text{CM}}(\mathcal{X},Q)$, but the large amount of numerical integration required to compute $C^{\text{CM}}(\mathcal{X})$ makes this intractable.  Instead we work with some well-established and more tractable lower bounds.

It is known that one can bound $C^{\text{CM}}(\mathcal{X})$ below by the \emph{cutoff rate} $R_0(\mathcal{X};H)$ \cite{hero}.  Following \cite{herath}, we can use the techniques of \cite{baccarelli} to compute the conditional cutoff rate given $H$ to be
\begin{eqnarray*}
R_0(\mathcal{X};H) &=& \log_2(M) \\
&-& \log_2\left[1+\frac{1}{M}\sum_{x\neq y\in H\mathcal{X}} \exp\left(\frac{-||x-y||^2}{4N_0}\right)\right]
\end{eqnarray*}
The cutoff rate is often used as a design criteria, and at low- and mid-range SNR constellations which optimize $R_0(\mathcal{X};H)$ are known to optimize $C^{\text{CM}}(\mathcal{X})$ as well \cite{baccarelli}.

To establish numerical techniques which are independent of $H$, the authors of \cite{herath} use Jensen's inequality on the random variable $H$ to establish a further lower bound $R_0(\mathcal{X};H)\geq R(\mathcal{X})$, where
\begin{equation*}
\boxed{ R(\mathcal{X}) = \log_2(M) - \log_2\left[1+\frac{1}{M}\sum_{x\neq y\in \mathcal{X}}\prod_{i=1}^n\frac{1}{1+\frac{|x_i-y_i|^2}{4N_0}}\right] }
\end{equation*}
which can be measured in bits/symbol.

In \cite{herath} the authors show that $R(\mathcal{X})$ is an excellent predictor of the behavior of $C^{\text{CM}}(\mathcal{X})$.  In particular, they show for low- and mid- range SNR values and $4$-QAM and $16$-QAM in $\C^4$ and $\C^6$, optimizing $R(\mathcal{X},Q)$ is essentially equivalent to optimizing $C^{\text{CM}}(\mathcal{X})$.  For the above reasons, we choose to only work with the objective function $R(\mathcal{X})$ in what follows, while keeping in mind the high correlation between $R(\mathcal{X})$ and $C^{\text{CM}}(\mathcal{X})$ established by the authors of \cite{herath}.

Our goal will be the following:
\begin{equation*}
   \boxed{ \text{Given $\mathcal{X}$ and SNR, compute $\arg\max_Q$ $R(\mathcal{X},Q)$. }}
\end{equation*}
One would like to construct rotation matrices for this purpose
\begin{itemize}
\item[$\bullet$] for all finite constellations $\mathcal{X}$ to allow for non-uniform constellations,
\item[$\bullet$] for all dimensions $n$,
\item[$\bullet$] which are independent of any parameterization of $n\times n$ rotation matrices, and
\item[$\bullet$] can be easily adapted to different levels of SNR.
\end{itemize}
Towards this end, we will construct for all $n = 2^k$ a family of rotations $Q_{2^k}(t)$ that offers excellent potential to maximize $R(\mathcal{X},Q)$ and therefore optimize the CM capacity.  The family $Q_{2^k}(t)$ depends on only one parameter $t\in [0,2\pi]$, and thus for a given $\mathcal{X}$ and SNR, the optimal rotation in the family can be easily computed by exhaustive search for any dimension.

\section{The Special Orthogonal Group $SO(n)$}\label{graddescent}

To properly describe the families of rotation matrices we construct, we need to discuss the structure of the \emph{special orthogonal group} $SO(n)$ of all $n$-dimensional rotation matrices.  The \emph{special orthogonal group} $SO(n)$ of rotations of $\R^n$ is defined by
\begin{equation*}
SO(n) = \{Q\in GL(n)\ |\ QQ^t = I_n,\ \det(Q) = 1\}
\end{equation*}
where $GL(n)$ is the group of all invertible real $n\times n$ matrices.  The dimension of $SO(n)$ as a manifold is $(n^2-n)/2$, which can be thought of as the minimum number of parameters required to describe an $n\times n$ rotation matrix.  We can now view our objective function $R(\mathcal{X},Q)$ as a function
\begin{equation*}
R(\mathcal{X},-):SO(n)\rightarrow \R
\end{equation*}
which measures the increase in CM capacity for varying rotation matrices.  The special orthogonal group is an example of a \emph{Lie group}, which is both a group and a manifold such that the group operations are continuous with respect to the manifold structure. As a general reference for the theory of Lie groups we recommend \cite{hallie}.  

The \emph{Lie algebra} $\frak{so}(n)$ of $SO(n)$ is the tangent space at the identity matrix $I_n\in SO(n)$, and thus is a real vector space of dimension $n(n-1)/2$.  We have the following convenient explicit description:
\begin{equation*}
\frak{so}(n) = \{A\in M(n)\ |\ -A = A^t\}
\end{equation*}
where $M(n)$ is the set of all $n\times n$ real matrices.  We pass from the Lie algebra to the Lie group using the exponential map $\exp:\frak{so}(n)\rightarrow SO(n),$ defined by the familiar power series
\begin{equation*}
\exp(A) = I_n + A + \frac{A^2}{2!} + \frac{A^3}{3!}+\cdots
\end{equation*}
One can verify easily that
\begin{equation*}
\exp(VAV^{-1}) = V\exp(A)V^{-1} 
\end{equation*}
for all $V\in GL(n)$.

A \emph{one-parameter subgroup} of $SO(n)$ is the image of a continuous group homomorphism $\R\rightarrow SO(n)$. One can show all one-parameter subgroups are of the form $Q(t) = \exp(At)$ for some $A\in\frak{so}(n)$ and $t\in \R$.  We can restate our goal using the language of Lie groups as follows:
\begin{equation*}
   \boxed{\parbox{.45\textwidth}{For fixed $n$, find a one-parameter family $Q_n(t)$ independent of $\mathcal{X}$ which approximates for varying SNR the local maxima of $R(\mathcal{X},-):SO(n)\rightarrow \R$.
}}
\end{equation*}
For the one parameter families we will construct it will suffice to restrict to $t\in [0,2\pi]$.  Thus for fixed $\mathcal{X}\subset\R^n$, and SNR, one can quickly compute the optimal $Q_n(t)$ by simple exhaustive search over a finite interval.

\section{Families of $2^k$-Dimensional Rotations}

In this section we will construct a family of candidates for good rotations for arbitrary constellations $\mathcal{X}\subset \R^{2^k}$ for any $k$.  First let us recall the definition of the Hadamard matrices $H_{2^k}\in M_{2^k}(\R)$:
\begin{equation*}
H_1 = [1],\ H_2 = \begin{bmatrix*}[r] 1 & 1 \\ 1 & -1 \end{bmatrix*},\ H_{2^k} = \begin{bmatrix*}[r] H_{2^{k-1}} & H_{2^{k-1}} \\ H_{2^{k-1}}  & -H_{2^{k-1}}  \end{bmatrix*}
\end{equation*}
We now construct skew-symmetric matrices $A_{2^k}\in \frak{so}(2^k)$ for $k\geq 1$ recursively in the following way:
\begin{equation*}
A'_1 = [0],\ A'_2 = \begin{bmatrix*}[r] A'_1 & H_1 \\ -H_1 & A'_1 \end{bmatrix*},\  A'_{2^k} = \begin{bmatrix*}[r] A'_{2^{k-1}} & H_{2^{k-1}}  \\ -H_{2^{k-1}}  & A'_{2^{k-1}} \end{bmatrix*}
\end{equation*}
\begin{equation*}
A_{2^k} = (2^k-1)^{-1/2}A'_{2^k}
\end{equation*}
The factor of $(2^k-1)^{-1/2}$ is only a convenience that simplifies some expressions in what follows.

Given a fixed $k$, we consider the one-parameter family of rotation matrices
\begin{equation}\label{master}
\boxed{Q_{2^k}(t) = \exp(A_{2^k}t)\in SO(2^k)}
\end{equation}
for $t\in \R$.  For a constellation $\mathcal{X}\subset \R^{2^k}$ and a fixed level of SNR, we can now compute the optimal $t\in[0,2\pi]$ which maximizes $R(\mathcal{X},Q_{2^k}(t))$ by simple exhaustive search.  Note that the family $Q_{2^k}(t)$ depends only on the dimension $n=2^k$, and not on $\mathcal{X}$.

The authors originally found the one-parameter family $Q_4(t)$ for 4D $4$-QAM using the geodesic flow algorithm (see \cite{karpukrot}) to numerically maximize the function $R(\mathcal{X},Q)$ over $SO(4)$.  For example, at $E_b/N_0 = 6$ dB geodesic flow on $SO(4)$ produced a matrix which was nearly identical to the matrix $Q_4(t)$ for $t = 0.8485$, the optimal $t$ for that level of SNR.  Similar results were obtained using the geodesic flow optimization method for other values of $E_b/N_0$ and other constellations, providing a large amount of experimental evidence suggesting $Q_{2^k}(t)$ contains a family of local maxima of the function $R(\mathcal{X},-):SO(n)\rightarrow \R$.  Our simulations also support this assertion, though the authors are currently unable to prove it.

It is worth noting that (the transpose of) the matrix $Q_4(0.5639)$ was considered in the DVB-NGH standard \cite{DVBNGH} to minimize the bit error rate at the demapper.  In a sense, one could describe our method as a generalization of the rotation matrix considered by DVB-NGH, but with the alternative purpose of increasing the CM capacity.

For general $n$ one of course could pick an arbitrary $A\in\frak{so}(n)$ and consider the one-parameter family $Q_n(t) = \exp(At)$, but preliminary simulations suggest that the performance of random one-parameter families is worse than the performance of the carefully constructed family $Q_{2^k}(t)$ above.

\section{Non-uniform QAM Constellations in $\R^2$}\label{nuqam2d}

Let us begin in $\R^2$.  The purpose of this section is two-fold: firstly, to demonstrate that simultaneous rotation and non-uniformity can improve the CM capacity of two-dimensional constellations, and secondly, to illustrate how optimal rotations can vary with constellations and the SNR.  We have $A_2 = \left[\begin{smallmatrix*}[r] 0 & 1 \\ -1 & 0 \end{smallmatrix*}\right]$ and hence the family $Q_2(t)$ is simply all of $SO(2)$:
\begin{equation*}
Q_2(t) = \exp(At) = \begin{bmatrix*}[r]
\cos(t) & \sin(t) \\ -\sin(t) & \cos(t)
\end{bmatrix*}
\end{equation*}

Let $M = 2^q$ with $q\geq 4$, and let $\alpha = (\alpha_1,\ldots,\alpha_{q-3})\in\R^{q-3}$.  By an \emph{$M$-NUQAM constellation} $\mathcal{X}(\alpha)\subset\R^2$ we mean the direct product of the set $\{\pm1,\pm\alpha_1,\ldots,\pm\alpha_{q-3}\}$ with itself.  For example, when $q=4$ and $\alpha = 4$, we have the $16$-NUQAM constellation
\begin{equation*}
\mathcal{X}(4) = \{(\pm1,\pm1),(\pm1,\pm4),(\pm4,\pm1),(\pm4,\pm4)\}.
\end{equation*}
Such constellations are known to increase CM capacity at lower SNR ranges, and $64$-NUQAM and $256$-NUQAM constellations have been included in the DVB-NGH standard \cite{DVBNGH}.  Unfortunately NUQAM constellations are not subsets of lattices, and thus the number theoretic techniques of \cite{OV}, which depend on the underlying lattice structure and large SNR are no longer applicable.

\begin{figure}[h!]
\includegraphics[width = .45\textwidth]{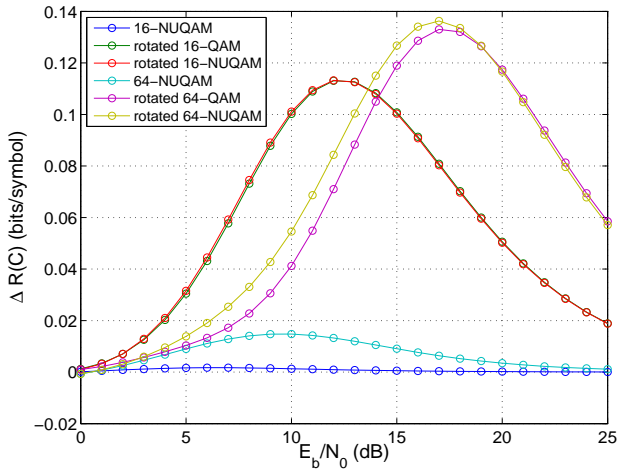}
\caption{Improvement in $R(\mathcal{X},Q_2(t))$ for optimally rotated $16$-NUQAM and $64$-NUQAM over the respective uniform, unrotated constellations.  For $16$-NUQAM, we selected $\alpha = 3.1903$, and for $64$-NUQAM, we selected $\alpha = (2.8727, 4.9280, 7.3827)$.}
\end{figure}

\begin{figure}[h!]
\includegraphics[width = .45\textwidth]{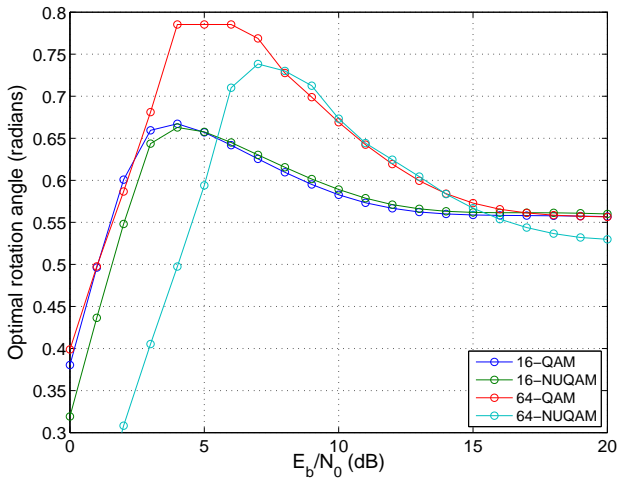}
\caption{Optimal rotations for some two-dimensional uniform and non-uniform constellations.}
\end{figure}

For a fixed $M$ and $E_b/N_0$, we consider the objective function $R(\mathcal{X}(\alpha))$, and perform gradient descent on the non-uniformity parameter $\alpha$.  We obtain the following:

\begin{center}
\begin{tabular}{|c|c|c|}
\hline
$E_b/N_0$ (dB) & $M$ & optimal $\alpha$ for $M$-NUQAM\\ \hline
8 & 16 & 3.1903 \\ \hline
12 & 64 & (2.8727, 4.9280, 7.3827) \\ \hline
\end{tabular}
\end{center}

To gauge the improvement of these non-uniform constellations over a large range of SNR values, we plot in Fig.\ 1 the function $R(\mathcal{X}(\alpha),Q)$ for rotated and unrotated versions of the above two $M$-NUQAM constellations.  At each value of $E_b/N_0$, we used the $t$ which maximizes $R(\mathcal{X},Q_2(t))$ for each constellation $\mathcal{X}$.

From Fig.\ 1 we can see that between 0 dB and 10 dB rotated $16$-NUQAM offers a slight improvement over $16$-NUQAM, while between 0 dB and 17 dB the effect of rotation on $64$-NUQAM is much more pronounced.  Thus combining non-uniformity and rotation does offer real benefits in terms of improving the CM capacity of a constellation.  From Fig.\ 2 we see that the per-SNR optimal rotation angles for each of the four constellations can be quite different, which underscores the necessity of finding optimal rotations specific to a given constellation and SNR.

\section{Rotated Constellations in $\R^4$}

Let us further examine the family (\ref{master}) for the simplest non-trivial example, that of rotated constellations in $\R^4$.  We have
\begin{equation*}
A_4 = \frac{1}{\sqrt{3}} \begin{bmatrix*}[r]
0 & -1 & 1 & 1 \\
1 & 0 & 1 & -1 \\
-1 & -1 & 0 & -1 \\
-1 & 1 & 1 & 0
\end{bmatrix*}
\end{equation*}
and for $t\in \R$, we can write $At = VDV^{-1}$ where
\begin{equation*}
D = \left[\begin{smallmatrix*}[r]
-it & & & \\
& -it & & \\
& & it &  \\
& & & it
\end{smallmatrix*}\right],\quad
V = \left[\begin{smallmatrix*}
\omega^2 & \omega & \omega^4 & \omega^5 \\
\omega & \omega^5 & \omega^5 & \omega \\
1 & 0 & 1 & 0 \\
0 & 1 & 0 & 1
\end{smallmatrix*}\right]
\end{equation*}
and $\omega = (1+\sqrt{-3})/2$.  It follows that
\begin{equation*}
Q_4(t) = V\exp(D)V^{-1} 
= \begin{bmatrix*}[r]
a & b & b & b \\
-b & a & b & -b \\
-b & -b & a & b \\
-b & b & -b & a
\end{bmatrix*}
\end{equation*}
where $a = \cos(t)$ and $b = \sin(t)/\sqrt{3}$.


\begin{figure}[h!]
\includegraphics[width = .45\textwidth]{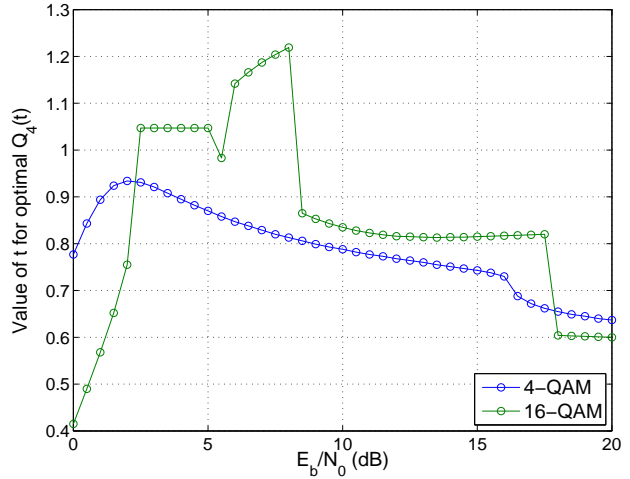}
\caption{Value of $t$ determining optimal rotation $Q_4(t)$ as a function of $E_b/N_0$, for both 4D $4$-QAM and $16$-QAM.}
\end{figure}

\begin{figure}[h!]
\includegraphics[width = .45\textwidth]{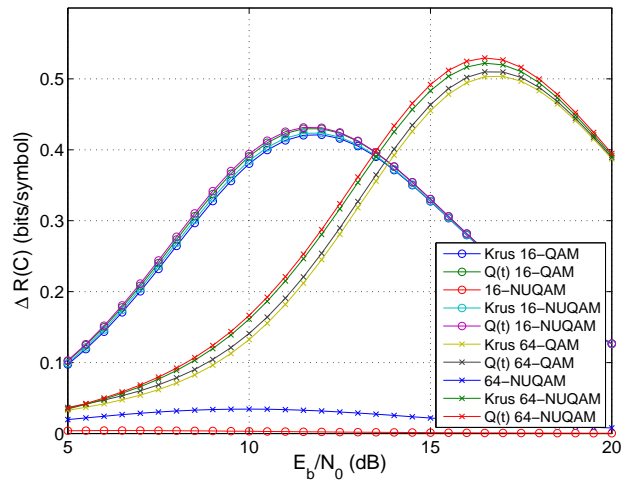}
\caption{Improvement in $R(\mathcal{X},Q_4(t))$ over unrotated 4D 16-NUQAM and 64-NUQAM constellations, using the optimal $t$ at each level of $E_b/N_0$.}
\end{figure}

For each level of $E_b/N_0$ we computed the optimal rotation $Q_4(t)$ for both 4D $4$-QAM and $16$-QAM, and plotted $t$ as a function of $E_b/N_0$ in Fig.\ 3.  This figure demonstrates that it is unlikely one can write $t$ as a 	``nice'' function of the SNR.  In Fig.\ 4 we plotted the increase in the function $R(\mathcal{X},Q)$ over unrotated $4$D $16$-QAM and $64$-QAM constellations, using our rotations for uniform and non-uniform constellations.    By 4D $M$-QAM or $M$-NUQAM constellation, we simply mean the direct product of two such 2D constellations.  The NUQAM constellations used were the respective direct products of the ones in the previous section.  We compared our rotation matrices to the Kr\"uskemper rotation in four dimensions (see the collection of algebraic rotations at \cite{Viterbo_rotations}), the optimal algebraic rotation in four dimensions.

\section{Rotated Constellations in $\R^8$}

Using similar methods as in the previous subsection, we can study the family (\ref{master}) when $k=3$, that is, rotations of constellations in $\R^8$.  We have
\begin{equation*}
Q_8(t) = \left[
\begin{smallmatrix*}[r]
a & b & b & b & b & b & b & b \\
-b & a & b & -b & b & -b & b & -b \\
-b & -b & a & b & b & b & -b & -b \\
-b & b & -b & a & b & -b & -b & b \\
-b & -b & -b & -b & a & b & b & b \\
-b & b & -b & b & -b & a & b & -b \\
-b & -b & b & b & -b & -b & a & b \\
-b & b & b & -b & -b & b & -b & a
\end{smallmatrix*} \right]
\end{equation*}
where $a = \cos(t)$ and $b = \sin(t)/\sqrt{7}$.  We collect the results of our simulations in Fig.\ 4, which compare $Q_8(t)$ to rotations constructed from totally real algebraic number fields (see \cite{Viterbo_rotations}).
\begin{figure}[h!]
\includegraphics[width = .45\textwidth]{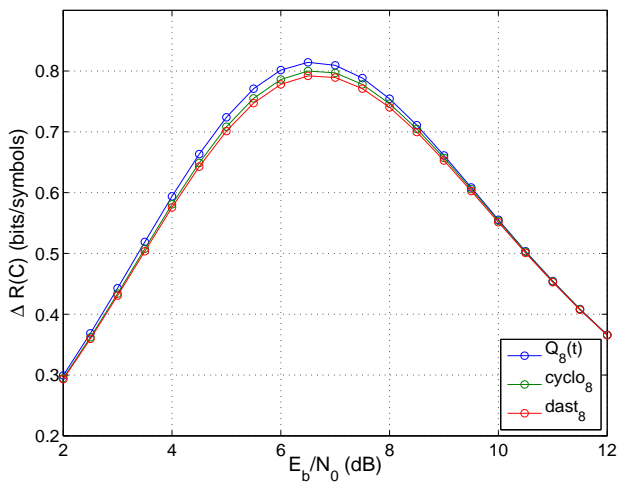}
\caption{Improvement in $R(\mathcal{X},Q_8(t))$ over unrotated 8D 4-QAM, using the optimal $t$ at each level of $E_b/N_0$.}
\end{figure}
One is tempted to compare our results to those obtained using the matrix $G_{\text{QPSK}}\in SO(4,\C)$ from \cite{herath}, but the comparison is unfair as the authors of \cite{herath} consider complex signal sets, and optimized $G_{\text{QPSK}}$ for a low value of $E_b/N_0$.  One can define a function $r:SO(4,\C)\rightarrow SO(8,\R)$ which maps each entry $a+bi$ to the matrix $\left[\begin{smallmatrix*}[r]a&-b\\ b&a\end{smallmatrix*} \right]$, but the resulting matrix $r(G_{\text{QPSK}})$ does not outperform any of the rotations used in Fig.\ 4.

\section{Conclusions and Future Work}\label{conclusions}

We have constructed a family of rotation matrices $Q_{2^k}(t)\in SO(2^k)$ for every $k$ with the goal of maximizing the CM capacity of arbitrary constellations in $\R^n$ at low- and mid-range SNR values.  Our approach is an adaptive per-SNR optimization, in which an optimal rotation for a given constellation and SNR value can be done by a simple exhaustive search on the interval $[0,\pi/2]$.  Our approach does not assume any structure in the constellation, and thus is applicable to non-uniform and uniform constellations alike, thereby improving the ability of non-uniform constellations to increase CM capacity.

We would like to extend our construction to arbitrary dimensions, not only those which are a power of two.  This could be done, for example, by considering block matrices whose blocks are of the form $Q_{2^k}(t)$ for varying $k$.  Furthermore, while numerical evidence and our simulations strongly support the implicit assertion that $Q_{2^k}(t)$ contains local maxima of the function $R(\mathcal{X},-):SO(n)\rightarrow \R$, one could potentially prove this by carefully analyzing the critical points of this function.

\section{Acknowledgements}

The first author has been partially supported by Academy of Finland grant 268364. Both authors have been partially supported by the Magnus Ehrnrooth Foundation, Finland.

\bibliographystyle{ieee}
\bibliography{myrefs_new}

\end{document}